# Wave-Driven Mixing Enhanced by Rotation in Red Giant Branch Stars

Simon Blouin[1], Paul R. Woodward[2], Pavel A. Denissenkov[1], Praneet Pathak[1], and Falk Herwig[1]

[1] Astronomy Research Centre and Department of Physics & Astronomy, University of Victoria, Victoria, BC, V8W 2Y2, Canada
[2] LCSE and Department of Astronomy, University of Minnesota, Minneapolis, MN 55455, USA

**Stars like our Sun expand as they exhaust their core hydrogen fuel, becoming red giants that eventually reach sizes up to 100 times their original radius. These giants have long presented a puzzle: they show systematic changes in their surface chemical composition that can only be explained by the transport of material from their nuclear-burning interior to their surface. The challenge is that this transport must somehow cross a stable layer that acts as a barrier between the star's outer convective envelope and its nuclear-burning interior. The convective motions in the envelope create internal waves that propagate through this barrier layer, but on their own these waves produce very little material transport. Here we show through high-resolution three-dimensional hydrodynamical simulations that stellar rotation dramatically amplifies how effectively these waves can mix material across this barrier. We find that the mixing rates can exceed those in non-rotating stars by over 100 times, increasing with faster rotation rates. This enhanced mixing provides a natural explanation for the observed chemical signatures in typical red giants. The amplification of wave-driven mixing by rotation may have implications beyond red giants to other types of stars.**

Internal waves arise naturally in density-stratified fluids, where displaced fluid parcels oscillate under the restoring force of buoyancy. In the Earth's oceans, these waves can drive mixing when they grow to sufficient amplitudes to break and generate turbulence [1,2]. Measuring wave-driven mixing in stars has proven more challenging. Observationally, their effects can only be detected indirectly, such as through changes in surface chemistry [3–5], making it difficult to unambiguously identify wave mixing as the cause. In simulations [6–9], wave mixing is typically weak and occurs over long timescales, making it challenging to measure reliably [10]. Artificial enhancements of stellar luminosities are often required to achieve numerically tractable fluid velocities [6,9,11–13], a procedure that makes it difficult to reliably extrapolate the measured mixing rates to realistic stellar conditions.

Our simulations tackle these numerical challenges using an explicit hydrodynamics code [9,14,15] that solves in three dimensions the Euler equations of compressible fluid dynamics using a high-order numerical method, leveraging massively parallel computing to achieve grids of up to $3456^3$. We focus on a 1.2 $M_\odot$ star approaching the tip of the red giant branch, where fluid velocities are naturally high enough so that artificial luminosity boosting is not necessary [6]. Our spherical-geometry computational domain captures the radiative barrier layer, extending from just above the hydrogen-burning shell into the inner 1-3% (by radius) of the convective envelope, with small-, medium-, and large-envelope simulations capturing progressively larger fractions of this region (see Fig. 1). Based on asteroseismic measurements of similar stars [16,17], we impose solid-body rotation at a rate of 0.7×10$^{-5}$ (slow rotating) to 7×10$^{-5}$ (fast rotating) radians per second. The simulations reveal a clear separation between turbulent convection in the envelope and wave-like motions in the interior, with the Coriolis force organizing the convective flows into columnar structures known as Taylor columns aligned with the rotation axis (Fig. 2).

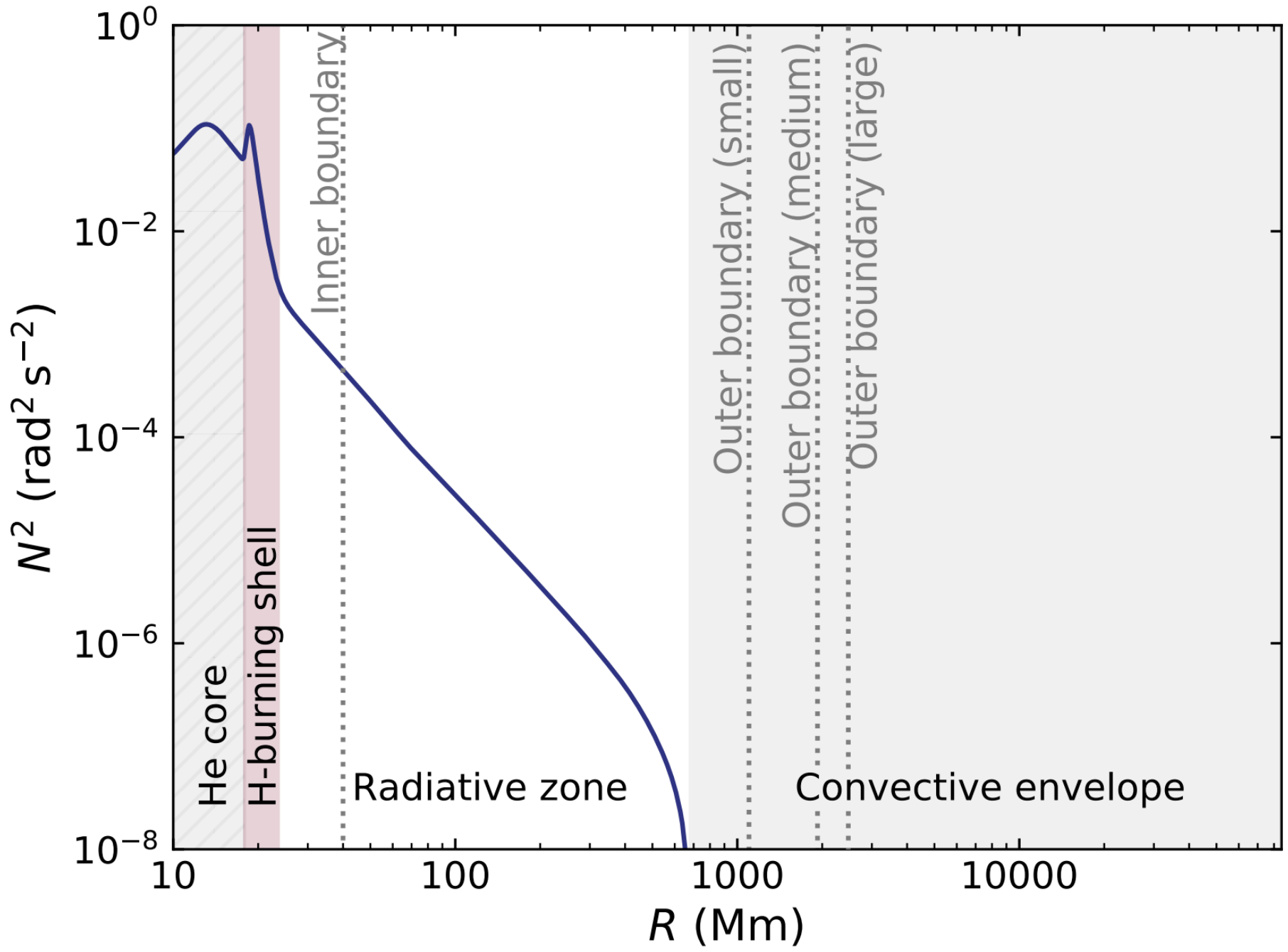

**Fig. 1 | Structure of our simulated red giant star.** The squared Brunt-Väisäla frequency ($N^2$) profile shows the internal stratification of our 1.2 $M_\odot$ red giant model. The simulation domain boundaries (vertical dotted lines) capture the radiative zone between the hydrogen-burning shell and the convective envelope, where we measure wave-driven mixing. The rightmost extent corresponds to the stellar surface, while shaded regions indicate the location of key structural features including the helium core, hydrogen-burning shell, and convective envelope.

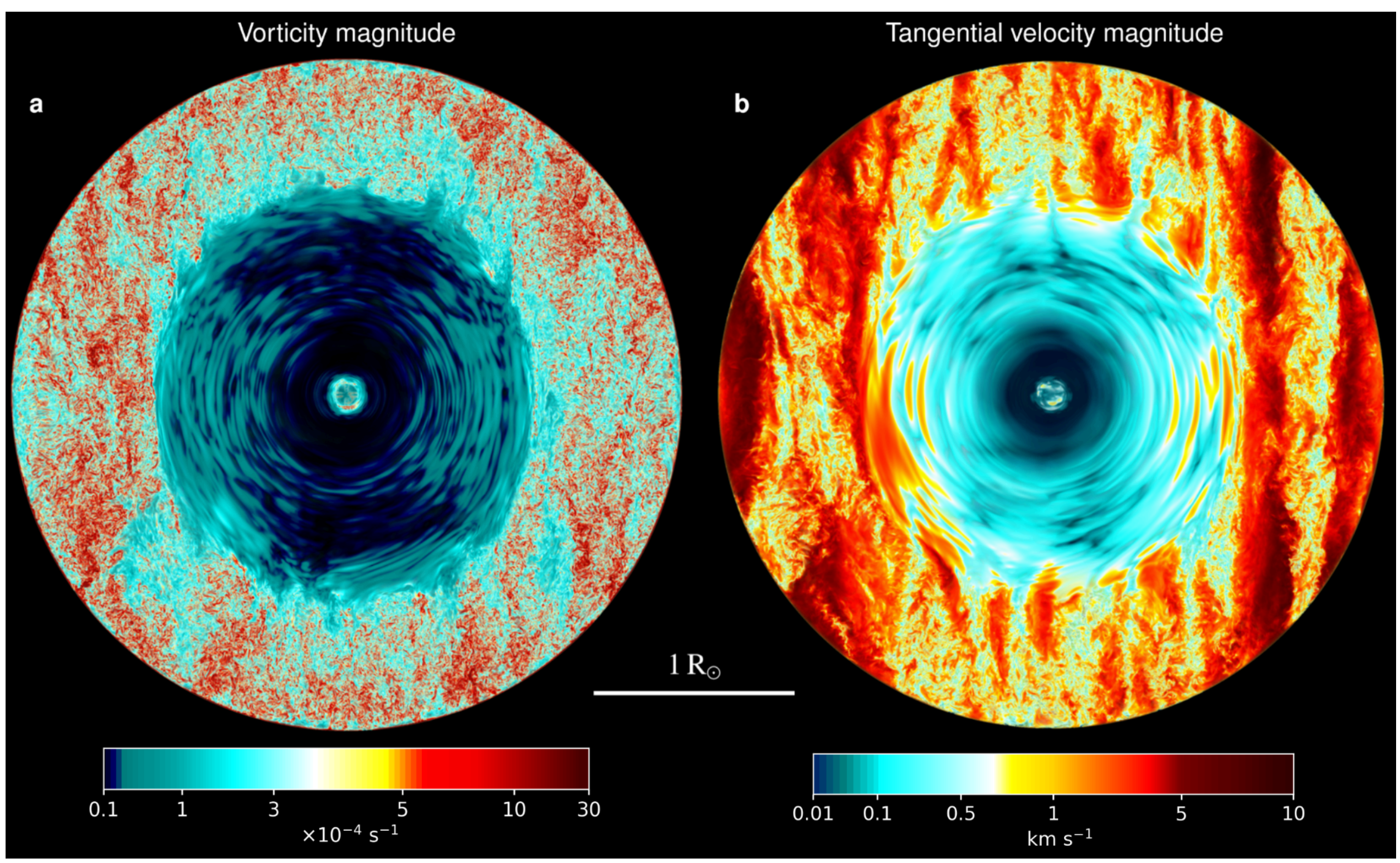

**Fig. 2 | Flow structure in a rotating red giant hydrodynamical simulation.** The images show a center-plane slice through the star's interior at $t$ = 646 h in our fast-rotating $2688^3$ small-envelope simulation, with the rotation axis oriented vertically. All quantities are displayed in the rotating reference frame. In panel a, the vorticity magnitude reveals turbulent motions in the outer convective envelope that contrast with the wave-dominated interior. In panel b, the tangential velocity magnitude highlights distinctive columnar structures aligned with the rotation axis, known as Taylor columns. The scale bar represents one solar radius. Artifacts are visible near the inner boundary (innermost 5% in radius) but do not affect the mixing region of interest [6]. The central blue region does not appear as an annulus because the finite-thickness slice (10% of the radial extent) captures the surface of the spherical inner boundary at 40 Mm, appearing as a filled circle. Individual grid layers within the slab are overplotted with partial transparency. The color scale is optimized for visual clarity of flow features rather than quantitative analysis. Movies are available at https://stellarhydro.ppmstar.org/ RGB-rotation/.

Our simulations reveal substantial mixing in the radiative zone, which we quantify by tracking the evolution of a passive tracer fluid initialized in spherical shells with Gaussian profiles (Fig. 3 and Methods). The mixing rates can be more than 100 times stronger than in equivalent non-rotating simulations [6], a result that persists when we reduce the grid size from $2688^3$ to a smaller $1536^3$ in the small-envelope setup, demonstrating numerical convergence. Additional simulations at different rotation rates (0.7, 2.2, and 7.0 × $10^{-5}$ rad/s) show that mixing efficiency increases strongly with rotation (Fig. 3).

When we extend our computational domain to include 3× and 4.5× the radial extent of the convective envelope compared to the small-envelope simulation shown in Fig. 2 (medium- and large-envelope simulations, respectively), the mixing becomes progressively more vigorous. We hypothesize that this enhancement occurs because larger envelope extents allow larger convective cells to develop (Extended Data Fig. 1), producing faster convective velocities that excite stronger waves [6]. While these measurements do not establish convergence with respect to domain size, they are consistent with a monotonic increase of mixing with modeled envelope extent toward an asymptote (Extended Data Fig. 2).

Under that monotonicity assumption, the measured diffusion coefficients are strict lower limits to the full-envelope values. Our interpretation would only be affected if the mixing were to decrease when including more of the envelope convection than we can currently model.

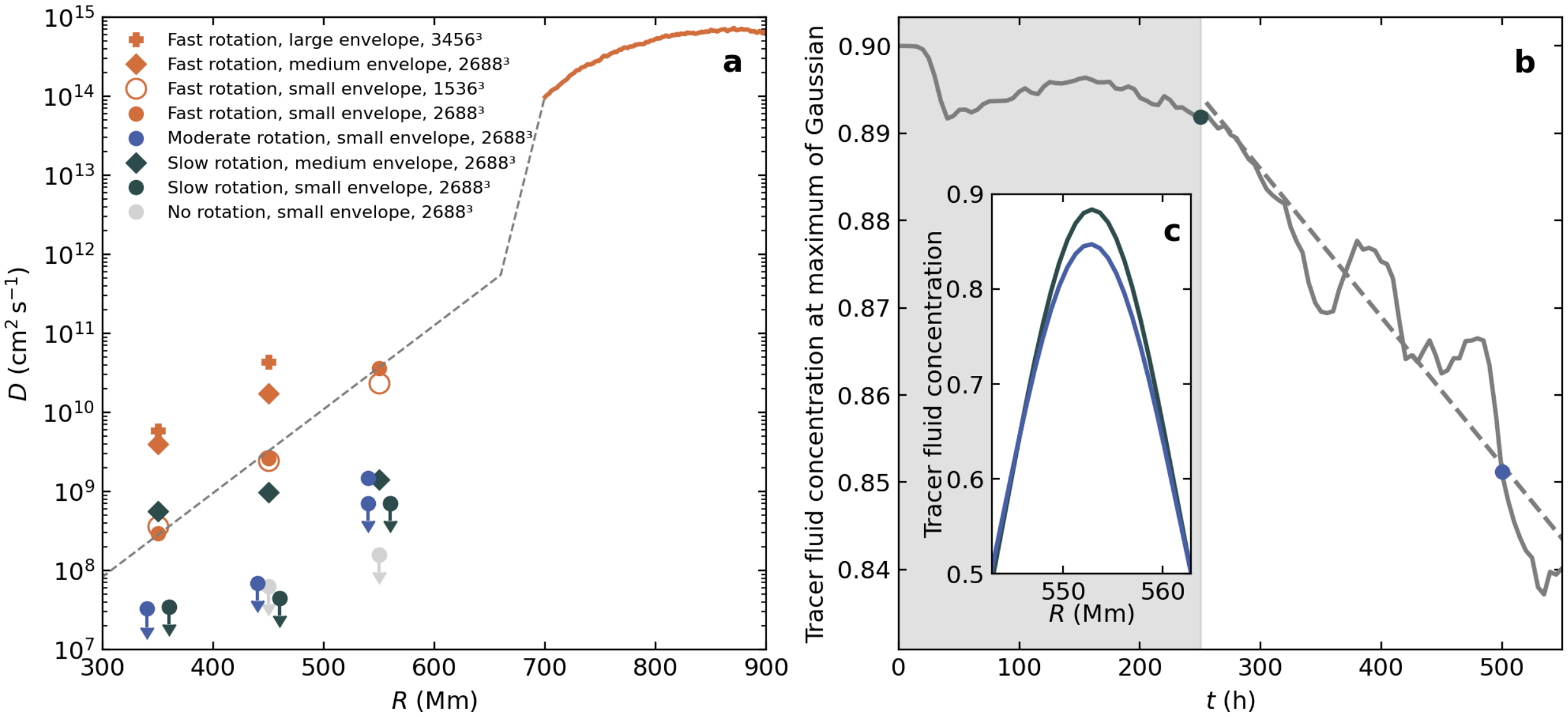

**Fig. 3 | Measurements of wave-driven mixing in the radiative zone.** We measured the diffusion coefficient $D$ at different radii in the radiative zone. In panel a, symbols show measurements obtained by tracking the spread of passive tracer fluid in spherical shells across different simulations. Some data points are horizontally offset by ±10 Mm from their nominal radii (350, 450, and 550 Mm) for visual clarity. The solid line shows $D$ obtained through inversion of tracer evolution [18] in the convection zone of the fast-rotating $2688^3$ small-envelope simulation, while the dashed line represents a fitted mixing model (see Methods). Results from a non-rotating simulation [6] are shown for comparison, demonstrating the dramatic enhancement of mixing with rotation. Panel b shows how we measured $D$ by tracking the peak amplitude of a Gaussian tracer shell over time (shown here for the fast-rotating $2688^3$ small-envelope simulation at 550 Mm). The dashed line shows the linear fit used to derive $D$, while the gray shaded area indicates the initial transient phase excluded from analysis. Panel c illustrates the spreading behavior quantified in panel b by showing radial profiles of a single Gaussian shell at two different times.

These dramatic rotation-induced enhancements in mixing emerge directly from solving the full fluid equations in three dimensions under realistic stellar conditions, without resorting to artificial luminosity boosting or simplified geometry. While our computational domain necessarily includes only a small fraction of the star's vast convective envelope, our medium- and large-envelope simulations suggest this limitation only leads us to underestimate the true mixing rates, provided mixing increases monotonically with modeled envelope extent. Previous work has found reduced wave mixing with rotation in 2D simulations of intermediate-mass main-sequence stars [19], but their geometry (core convection) and dimensionality (no Taylor columns possible) differ fundamentally from our 3D simulations. The Taylor columns create strong departures from spherical symmetry in both the convective flow patterns and their interaction with the radiative zone (see Methods), in contrast to non-rotating convection which is spherically symmetric when time-averaged. Understanding the detailed physics of how rotation modifies wave properties [19–21] and enhances mixing remains an open question, but the strong increase in mixing rates is a robust feature of our simulations.

To explore the implications for observed stellar abundance predictions, we modeled the one-dimensional stellar evolution of a 1.2 $M_\odot$ star incorporating our measured mixing rates (Fig. 4). Since we can only directly simulate conditions near the tip of the red giant branch where fluid velocities are high enough, we tested two scenarios: one using constant mixing rates from our fitted model (dashed line in Fig. 3) throughout the upper red giant branch, and another where that radial profile is scaled linearly with luminosity along the RGB (reaching the full values only at the RGB tip) and increased uniformly by 1 dex. This enhancement is motivated by the ~1 dex increase measured between the small-envelope runs (to which the mixing model was fit) and the medium/large-envelope runs. Both scenarios successfully reproduce the observed decline in carbon isotope ratios, demonstrating that rotationally-enhanced wave mixing provides a viable solution to this long-standing puzzle [22–26].

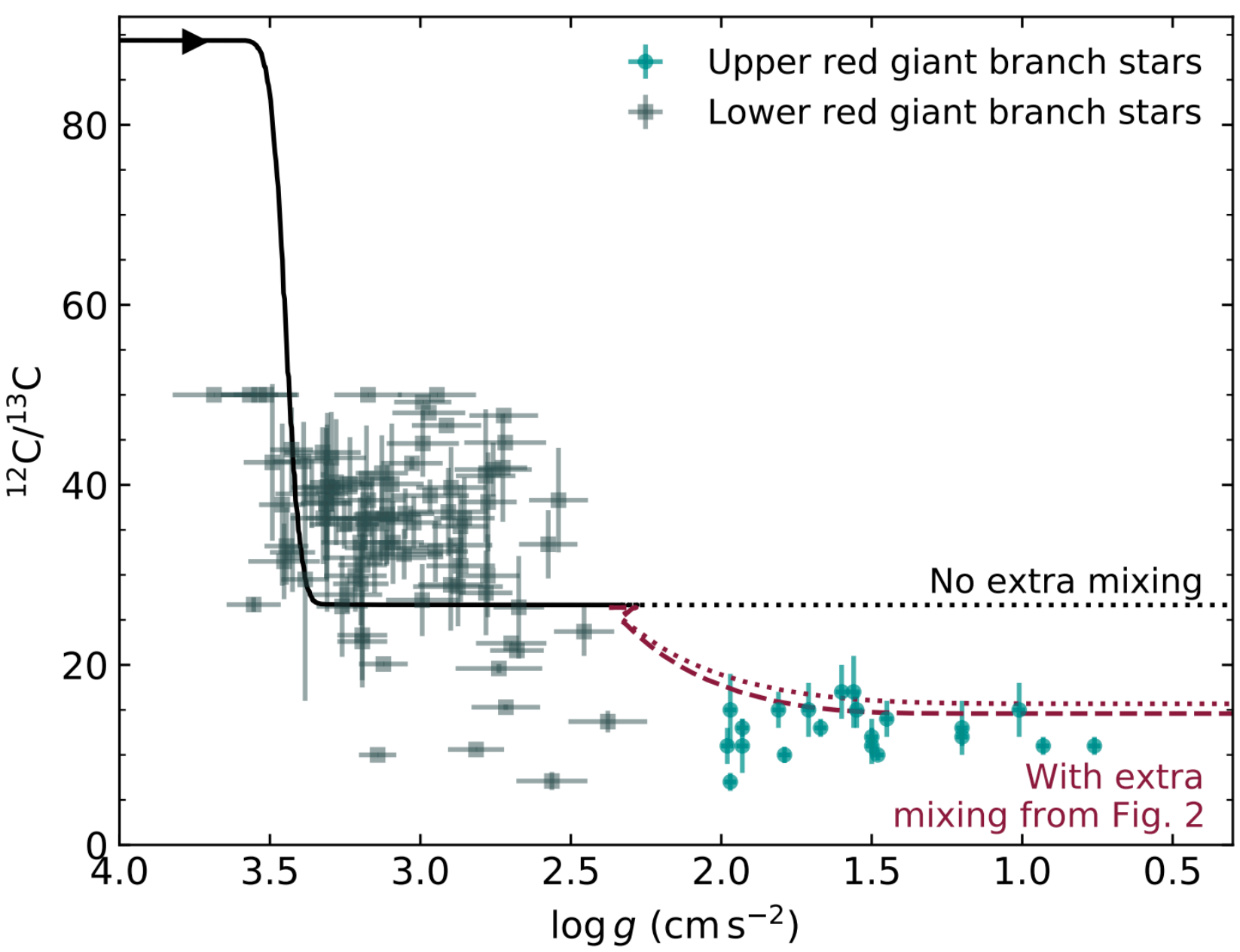

**Fig. 4 | Rotation-enhanced wave mixing and surface chemical changes in red giant stars.** The evolution of carbon isotope ratios is shown for a 1.2 $M_\odot$ star, starting from its initial solar-like value ($^{12}C/^{13}C = 89$) on the left. The sharp decline at a surface gravity log $g = 3.5$ is the result of a well-understood convective process that accompanies the transformation of a solar-like star into a red giant. The black line shows a model without extra mixing, which maintains a constant $^{12}C/^{13}C$ ratio after this initial decline, failing to match the observed continued decline along the upper red giant branch. Models incorporating our simulated wave mixing (red lines) successfully match the observations, whether using constant mixing rates from our fitted model (dashed line) or a diffusion coefficient profile 1 dex higher (to approximate the full-envelope effect) that is scaled linearly with luminosity along the RGB to reach its full values only at the tip (dotted line). The colored points show observational data from stars of similar masses and metallicities ($M < 2$ $M_\odot$ and $-0.4 \leq$ [Fe/H] $\leq 0.3$) [27,28].

The steep dependence we found between mixing efficiency and rotation suggests that rotation-enhanced mixing could potentially reach levels sufficient to enable lithium production through the Cameron–Fowler mechanism [26,29] in the most rapidly rotating giants. This possibility is particularly interesting given recent surveys that have revealed a population of red giants with unexpectedly high lithium abundances [30,31], and observations that these lithium-rich giants are preferentially fast rotators [32,33]. The finding that rotation can so strongly amplify wave-driven mixing opens new possibilities for understanding chemical transport in stars, as this mechanism likely operates in other phases of stellar evolution where waves and rotation coexist.

## Methods

**Initial stellar structure.** We modeled a 1.2 $M_\odot$ star near the tip of the red giant branch using the MESA stellar evolution code [6,34–36]. The initial stratification was mapped onto a 3D Cartesian grid in the PPMstar hydrodynamics code [9,14], which solves the fluid equations in $4\pi$ geometry with radiation diffusion [15] and realistic opacities [6,15]. These simulations function as implicit large eddy simulations, relying on the numerical properties of PPMstar's scheme to provide dissipation at the grid scale [37].

The computational domain captures the radiative region between the hydrogen-burning shell and the convective envelope (Fig. 1). The inner boundary at 40 Mm avoids the electron degenerate core, as the current PPMstar equation of state only includes contributions from ideal gas and radiation pressure [15]. The outer boundary was set at either 1100, 1925, or 2475 Mm (for our small-, medium-, and large-envelope simulations, respectively), corresponding to 1 to 3% of the total stellar radius. While this captures only a small fraction of the extensive convective envelope, it includes the key region where mixing between the nuclear-burning interior and envelope should occur. However, the convective motions that excite internal waves depend on non-local properties of the convection zone, meaning that the extent of the envelope we include can affect our results. Our extended-envelope simulations directly address this limitation by allowing us to evaluate how the measured mixing rates depend on the extent of the simulated envelope. Reflective boundary conditions were applied at both the inner and outer radial boundaries, with gravity linearly turned off with radius starting in the last ~1% of radial extent near the boundaries.

We performed simulations at three resolutions, $1536^3$, $2688^3$ and $3456^3$ grid cells, running for approximately 30 days of stellar evolution time in each case (Supplementary Table 1). All simulations used the star's nominal luminosity of 2100 $L_\odot$. Note that the $3456^3$ large-envelope, $2688^3$ medium-envelope and $1536^3$ small-envelope simulations all maintain the same grid cell size.

**Rotation implementation.** We chose angular velocities in the range $\Omega = 0.7 \times 10^{-5}$ to $7 \times 10^{-5}$ rad/s. While RGB stars rotate differentially with faster-spinning cores [17,38], our simulations assume solid-body rotation at a rate representative of the radiative zone where wave mixing occurs. Earlier RGB stars with radii of 5–7 $R_\odot$ have envelope rotation frequencies in the $(2–6) \times 10^{-7}$ rad/s range [16]. Assuming conservation of specific angular momentum ($R^2\Omega$), $\Omega = 7 \times 10^{-5}$ rad/s represents a plausible rotation frequency at the center of the radiative zone ($R \approx 0.5\ R_\odot = 348$ Mm), while asteroseismic constraints on central rotation rates in RGB stars [17] suggest that $\Omega \approx 1 \times 10^{-5}$ rad/s may be more typical. The simulations were performed in a rotating reference frame, with the Coriolis and centrifugal forces explicitly calculated. To minimize transient effects, we gradually accelerated the coordinate frame over $t_0 = 150{,}000$ s according to $\Omega(t) = (\Omega_0/2)[1 + \sin(\pi(t - t_0/2)/t_0)]$ for $t < t_0$, after which $\Omega = \Omega_0$. Our analysis considers only the portion of the simulations after at least $6t_0 = 250$ h, when the gas dynamics is well established.

The simulations exhibit a gradual increase in convective and wave velocities over time that is not seen in non-rotating runs [6]. For example, convective velocities in the rotating frame increase by about one-third during the analysis period for our fast-rotating $2688^3$ small-envelope simulation. We attribute this drift to imperfect angular momentum conservation, which is maintained to within 1%. The fluid velocities in the rotating frame of reference are much smaller than the reference frame rotation velocity. Convective velocities are of the order of 3 km/s, while the bulk rotation velocity in the envelope reaches 70 km/s for $\Omega = 7 \times 10^{-5}$ rad/s. Therefore, even small angular momentum conservation errors can noticeably affect the fluid velocities in the rotating frame. While we observe a similar drift in velocities in the stable layers, the measured mixing rates remain constant on average throughout our simulations. Our diffusion coefficient measurements show no systematic increase with time.

**Flow patterns.** The Coriolis force in our rotating reference frame organizes the convective motions into a distinctive pattern of counter-rotating flows (Supplementary Figure 1). In the rotating frame, we observe retrograde flow at the bottom of the convective envelope and prograde flow at the top, though all motions are prograde in the inertial frame but with different angular velocities. The convective penetration beyond the Schwarzschild boundary shows strong latitudinal variation (Supplementary Figure 2). Convective motions extend significantly beyond this boundary near the equator while remaining more sharply confined at the poles. This is visible in both the velocity fields and temperature gradient profiles.

**Mixing rate measurements.** We quantified mixing in the radiative zone by tracking the evolution of a passive tracer fluid using PPMstar's high-order piecewise parabolic Boltzmann (PPB) advection scheme [14]. The tracer fluid was initialized in spherical shells with Gaussian radial profiles at three different radii (350, 450, and 550 Mm) in the radiative zone. Under diffusive mixing, these profiles spread while maintaining their Gaussian shape, allowing us to derive a diffusion coefficient $D$ from the rate at which the peak amplitude decreases:

$$D = - \sigma_0^2 \, (\mathrm{d}FV_{max}/\mathrm{d}t)/FV_{max,0}$$

where $FV_{max}$ is the tracer concentration at the maximum of the Gaussian, $FV_{max,0}$ is its initial value, and $\sigma_0$ is the initial standard deviation. This direct measurement approach avoids the inaccuracies associated with post-processing using tracer particles. This measured diffusion coefficient quantifies the net mixing produced by the resolved dynamics together with the implicit subgrid dissipation of our implicit large-eddy simulations.

We omit measurements at 550 Mm in the medium- and large-envelope simulations because stronger convective motions lead to convective intrusions that overlap with the 550 Mm Gaussian, making it impossible to isolate and measure the wave-driven mixing signal from the tracer evolution.

**Evolution of surface abundances.** We modeled the long-term impact of wave-driven mixing using MESA (version 7624). The calculations followed a 1.2 $M_\odot$ star with [Fe/H] = -0.3 from the main sequence through the red giant branch. We implemented extra mixing using a double-*f* prescription [39], where the diffusion coefficient at a distance $z$ below the convective boundary is given by:

$$D(z) = D_0 \exp[-2z/(f_1 H_P)] \text{ for } z \leq z_2 \,,$$

$$D(z) = D_2 \exp[-2(z-z_2)/(f_2 H_P)] \text{ for } z > z_2$$

where $D_2 = D_0 \exp[-2z_2/(f_1 H_P)]$ ensures continuity at $z_2$= 40 Mm [6]. Here, $H_P$ is the local pressure scale height, and $f_1$ and $f_2$ are free parameters controlling the mixing length scales. The mixing was restricted to a maximum depth of 0.055 $R_\odot$ (38 Mm) above the hydrogen-burning shell, consistent with theoretical constraints [26]. We calibrated $D_0$ to match our hydrodynamical simulation measurements, testing two scenarios: one with constant mixing throughout the upper RGB using $D_0 = 10^{14}$ cm²/s (dashed line in Fig. 3), and another where we adopt a diffusion coefficient profile 1 dex higher (to approximate the full-envelope effect) but scale this entire profile linearly with luminosity along the RGB, reaching these full values only at the tip. For both scenarios, we used $f_1$ = 0.06 and $f_2$ = 0.32 to reproduce the radial profile of mixing measured in our simulations (Fig. 3 and ref. [6]).

## Data availability

A representative sample of radial profiles and filtered 3D output data from the PPMstar simulations is available on Figshare at https://doi.org/10.6084/m9.figshare.30273592. The PPMstar outputs can also be explored on ppmstar.org.

## Code availability

PPMstar is a proprietary code and is not publicly available. MESA is available online at https://sourceforge.net/projects/mesa/files/releases/mesa-r7624.zip. Python tools used for the data analysis in this study are publicly available on GitHub at https://github.com/PPMstar/PyPPM. Example notebooks can be found at https://github.com/PPMstar/PPMnotebooks.

## Acknowledgments

We thank Janosz Dewberry for helpful discussions on the impact of rotation on waves. Claude Sonnet 3.5 and Opus 4.0 were used to improve wording at the sentence level and assist with coding. F.H. acknowledges funding through an NSERC Discovery Grant. P.R.W. acknowledges funding through NSF CDS&E grants 1814181 and 2309101. This work benefited from support by the U.S. Department of Energy, Office of Science, Office of Nuclear Physics, under Award Number DE-SC0023128 (CeNAM). The simulations for this work were carried out on the NSF Frontera supercomputer operated by the Texas Advanced Computing Center at the University of Texas at Austin and on the Digital Research Alliance of Canada Niagara and Trillium supercomputers operated by SciNet at the University of Toronto. The data analysis was carried on the Astrohub online virtual research environment (https://astrohub.uvic.ca) developed and operated by the Computational Stellar Astrophysics group (http://csa.phys.uvic.ca) at the University of Victoria and hosted on the Digital Research Alliance of Canada Arbutus Cloud operated by Research Computing Services at the University of Victoria.

## Author contributions

S.B., P.A.D. and F.H. planned this project and wrote the manuscript. S.B. set up and analyzed the hydrodynamical simulations. P.R.W. developed and maintained the PPMstar code used to perform these simulations, and provided guidance on its optimal use. P.A.D. performed the stellar evolution calculations and provided expertise on stellar evolution and observational constraints. S.B., P.R.W., F.H., and P.P. developed numerical tools used to analyze the outputs of the simulations. All authors contributed to the interpretation of the results and the refinement of the manuscript.

## Competing interests

The authors declare no competing interests.

## Extended Data

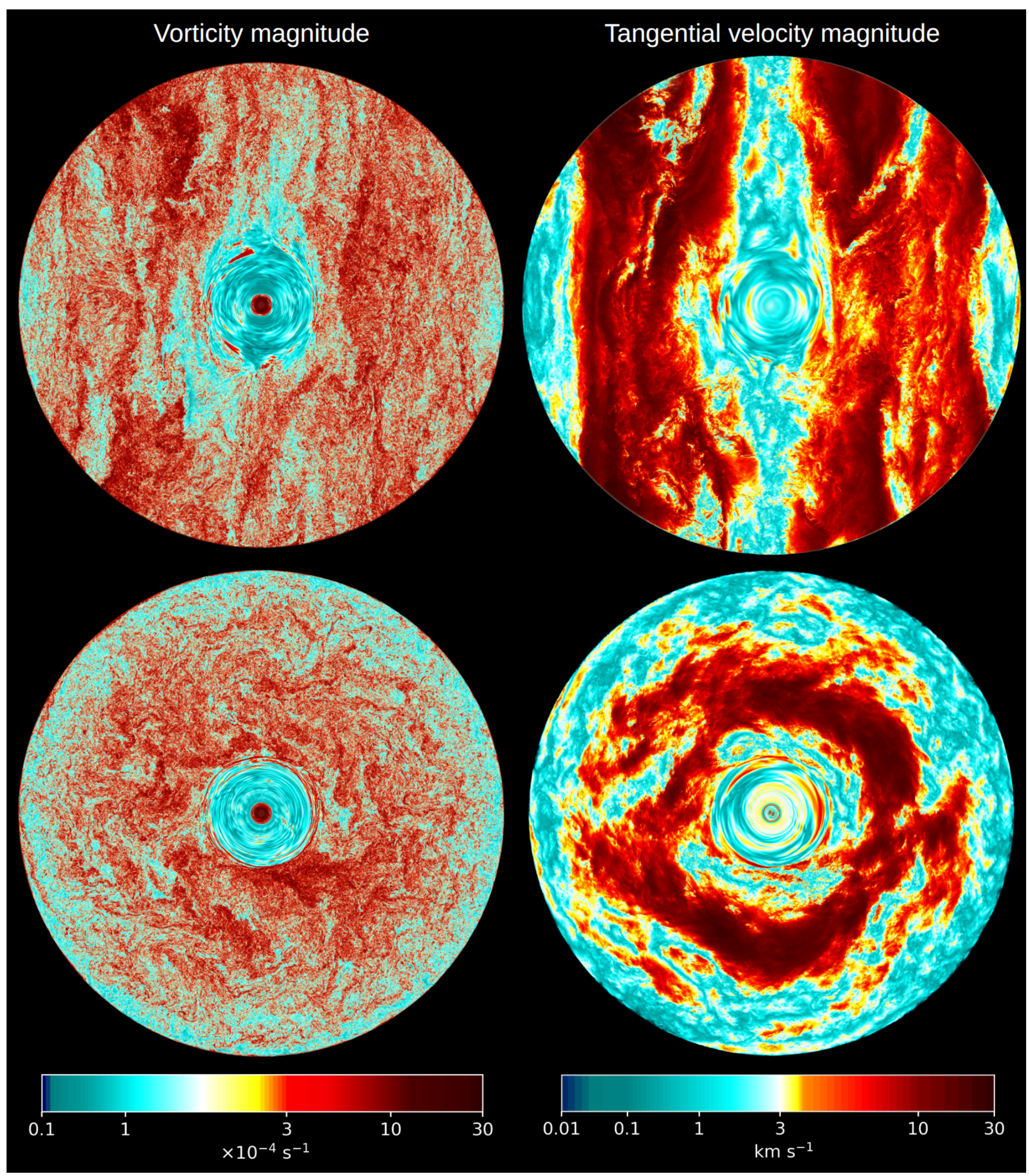

**Extended Data Figure 1 | Flow structure in the large-envelope rotating red giant simulation.** The images show center-plane slices through the star's interior at $t$ = 739 h in our fast-rotating $3456^3$ large-envelope simulation. The top row shows the equator-on view with the rotation axis oriented vertically, while the bottom row shows the pole-on view. All quantities are displayed in the rotating reference frame. The extended radial domain (reaching 2475 Mm) captures a larger fraction of the convective envelope compared to Fig. 2, allowing more developed structures in the convection zone.

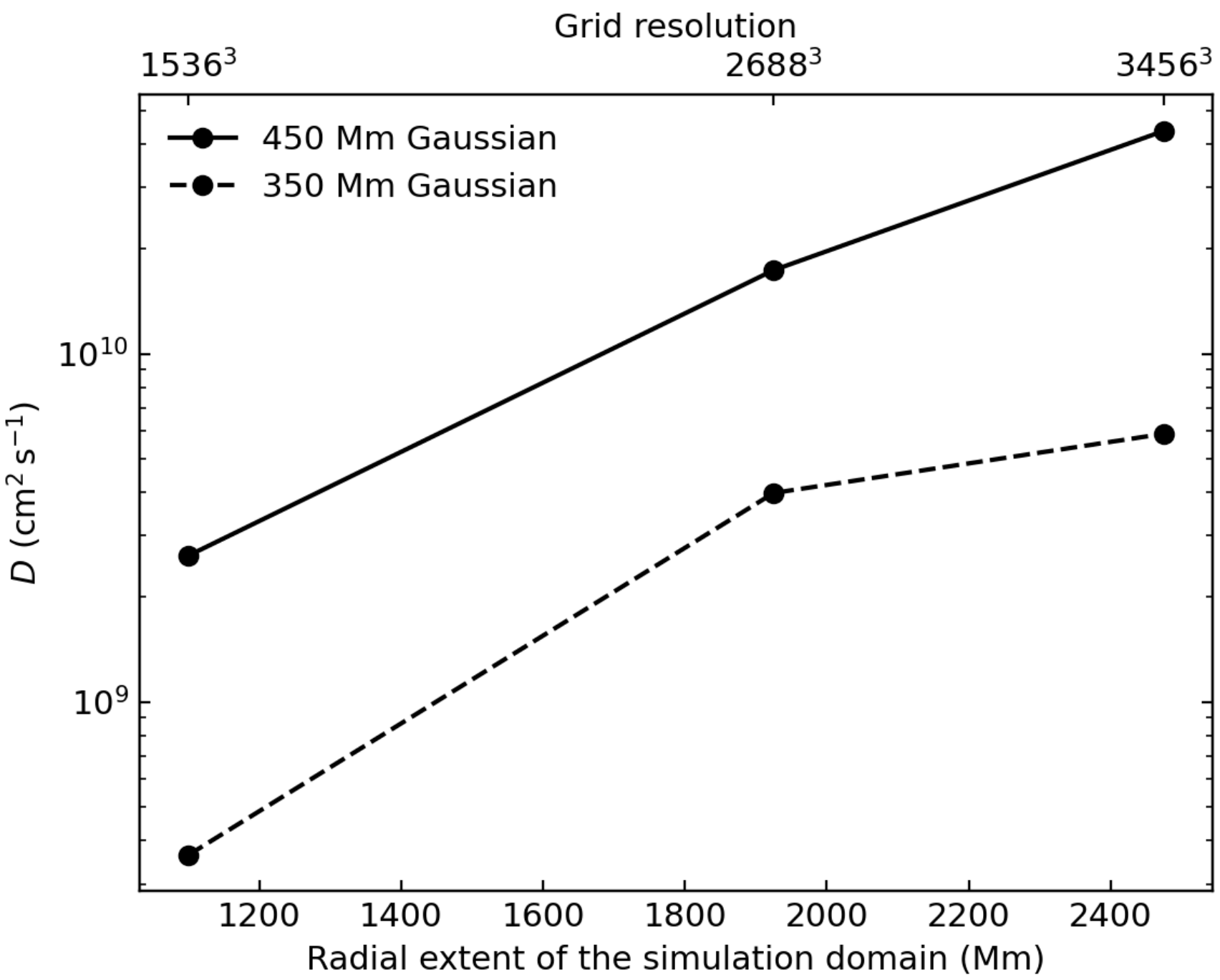

**Extended Data Figure 2 | Convergence of mixing measurements with domain size.** The diffusion coefficient $D$ measured at two radii in the radiative zone (350 Mm and 450 Mm Gaussian tracers) as a function of the radial extent of the computational domain. All simulations shown are performed at $\Omega = 7 \times 10^{-5}$ rad/s and with identical grid cell sizes. The mixing rates increase with larger envelope extents, indicating that our simulations have not yet reached full convergence with respect to domain size. However, the trend suggests we are approaching convergence, with the rate of increase diminishing at larger envelope extents. Measurements at 550 Mm are excluded due to convective intrusions that prevent unambiguous isolation of the wave-driven mixing signal.

| Run ID | Grid resolution | Simulation domain extent (Mm) | Rotation Rate ($\times 10^{-5}$ rad/s) | Total time (days) |
|---|---|---|---|---|
| X39 | $2688^3$ | 1100 | 7.0 | 23.6 |
| X38 | $1536^3$ | 1100 | 7.0 | 47.2 |
| X49 | $2688^3$ | 1925 | 7.0 | 42.0 |
| X78 | $3456^3$ | 2475 | 7.0 | 30.8 |
| X52 | $2688^3$ | 1100 | 2.2 | 24.5 |
| X51 | $2688^3$ | 1100 | 0.7 | 23.2 |
| X82 | $2688^3$ | 1925 | 0.7 | 57.6 |

**Supplementary Table 1 | Simulation parameters.** Overview of the three-dimensional hydrodynamical simulations performed in this work. For reference, the convective turnover timescale, estimated as the convective zone's radial extent within the domain divided by the mean RMS convective velocity, is ~2 days.

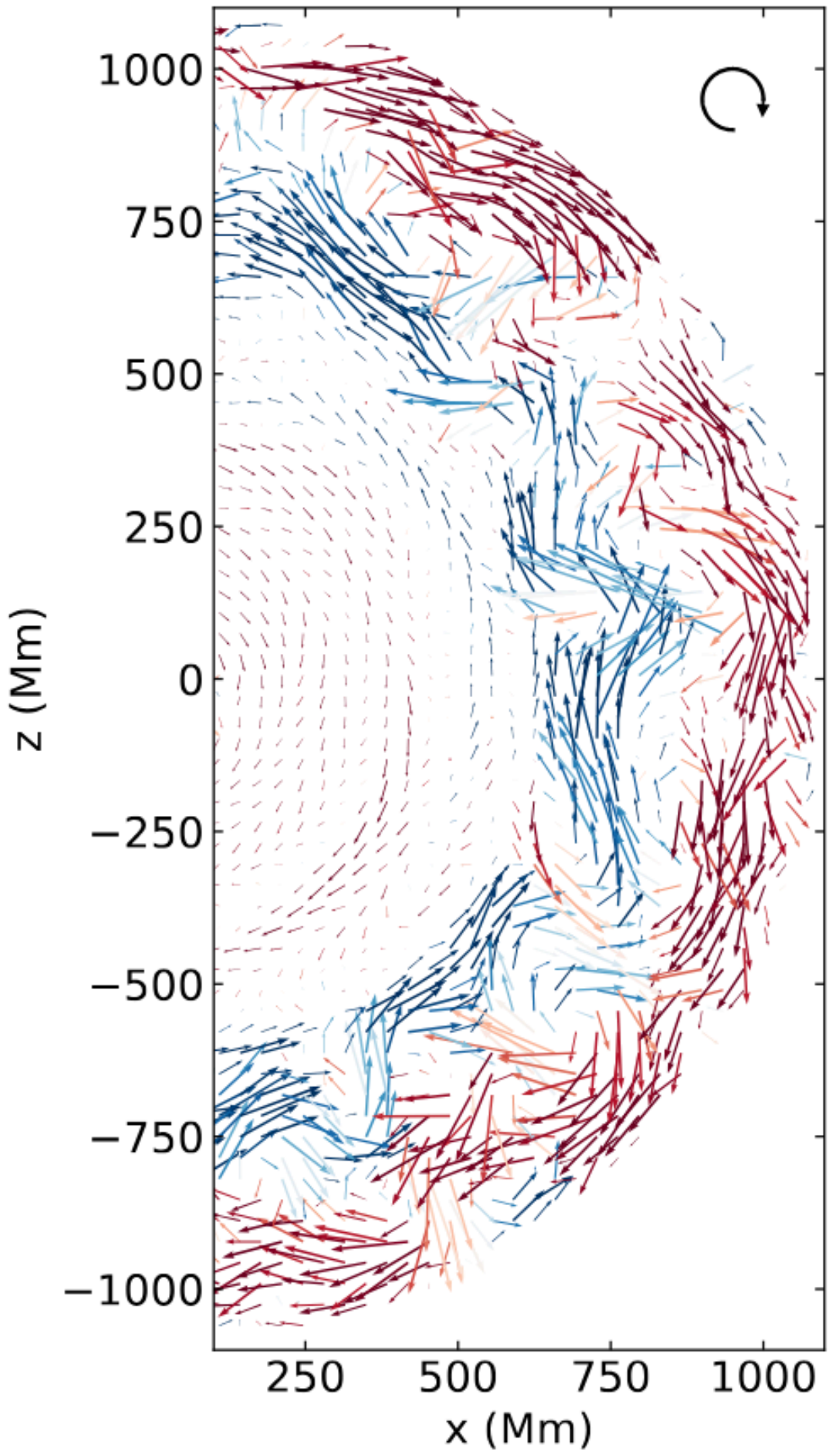

**Supplementary Figure 1 | Counter-rotating flows in the rotating reference frame.** The velocity field is shown in a center-plane slice of our $1536^3$ simulation and prepared from filtered 3D output data [40], with the rotation axis perpendicular to the page. The Coriolis force organizes convective motions into distinct flow patterns. Red and blue colors indicate flow moving with (prograde) and against (retrograde) the stellar rotation direction respectively, while arrows show the local flow direction and magnitude. In the inertial frame, all motions are prograde but with different angular velocities that create this organized pattern in the rotating frame.

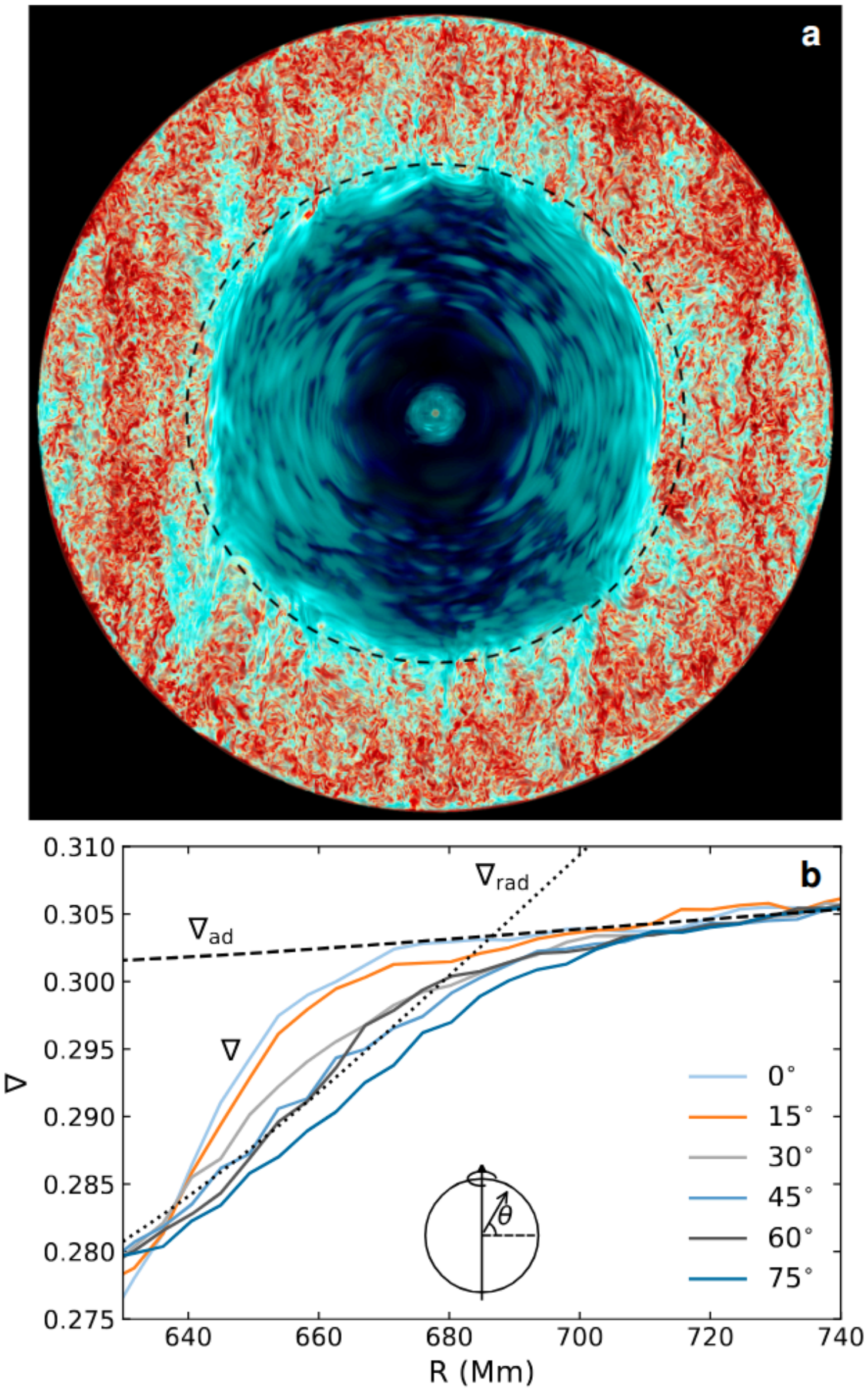

**Supplementary Figure 2 | Latitude-dependent convective penetration.** The rotation in our simulations creates a strong asymmetry in how convective motions penetrate beyond the Schwarzschild boundary. In panel a, a side-view of vorticity magnitude (with rotation axis oriented vertically) in our fast-rotating small-envelope simulation at $t$ = 410 h shows how convective motions extend significantly beyond the Schwarzschild boundary (dashed circle) near the equator while remaining confined at the poles. Panel b quantifies this asymmetry through temperature gradients measured at different latitudes θ, compared to spherically averaged adiabatic ($\nabla_{ad}$) and radiative ($\nabla_{rad}$) gradients. The temperature stratification near the equator remains adiabatic beyond the Schwarzschild boundary, indicating the establishment of a convective penetration zone at these latitudes.